\documentclass[]{aastex701}
\usepackage{color}
\usepackage{xcolor}

\begin{document}

\title{SN 2023gfo: A Peculiar Type IIP Supernova with High Luminosity and Normal Plateau Duration

}

\author{Riko Namba}
\affiliation{Graduate School of Science and Engineering, Kagoshima University, 1-21-35 Korimoto,Kagoshima, Kagoshima 890-0065, Japan}
 \affiliation{Department of Physics and Astronomy, Faculty of Science, Kagoshima University, 1-21-35 Korimoto, Kagoshima, Kagoshima 890-0065, Japan}
\email{k3015064@kadai.jp}
\author{Masayuki Yamanaka}
\affiliation{Amanogawa Galaxy Astronomy Research Center (AGARC), Graduate School of Science and Engineering, Kagoshima University, 1-21-35 Korimoto, Kagoshima, Kagoshima
 890-0065, Japan}
\email{yamanaka@sci.kagoshima-u.ac.jp}

\author{Keiichi Maeda}
\affiliation{Department of Astronomy, Kyoto University, Kitashirakawa-Oiwake-cho, Sakyo-ku, Kyoto 606-8502, Japan}
\email{keiichi.maeda@kusastro.kyoto-u.ac.jp}

 \author{Avinash Singh}
 \affiliation{Oskar Klein Centre, Department of Astronomy, Stockholm University, AlbaNova, SE-106 91 Stockholm, Sweden}
 \email{avinash21292@gmail.com}

\author{Kenta Taguchi}
\affiliation{Okayama Observatory, Kyoto University, 3037-5 Honjo, Kamogatacho, Asakuchi, Okayama 719-0232, Japan}
\affiliation{Department of Astronomy, Kyoto University, Kitashirakawa-Oiwake-cho, Sakyo-ku, Kyoto 606-8502, Japan}
\email{kentagch@kusastro.kyoto-u.ac.jp}

\author{Takahiro Nagayama}
\affiliation{Graduate School of Science and Engineering, Kagoshima University, 1-21-35 Korimoto,Kagoshima, Kagoshima 890-0065, Japan}
\email{nagayama@sci.kagoshima-u.ac.jp}

\author{Miho Kawabata}
\affiliation{Okayama Observatory, Kyoto University, 3037-5 Honjo, Kamogatacho, Asakuchi, Okayama 719-0232, Japan}
\email{kawabata@kusastro.kyoto-u.ac.jp}

\author{Koji S. Kawabata}
\affiliation{Hiroshima Astrophysical Science Center, Hiroshima University, 1-3-1 Kagamiyama, Higashi-Hiroshima, Hiroshima 739-8526, Japan}
\email{kawabtkj@hiroshima-u.ac.jp}

 \author{Tatsuya Nakaoka}
\affiliation{Hiroshima Astrophysical Science Center, Hiroshima University, 1-3-1 Kagamiyama, Higashi-Hiroshima, Hiroshima 739-8526, Japan}
\email{nakaokat@hiroshima-u.ac.jp}

\author{Devendra Sahu}
\affiliation{Indian Institute of Astrophysics, Koramangala 2nd Block, Bangalore 560034, India}
\email{dks@iiap.res.in}

\author{Anjasha Gangopadhyay}
\affiliation{Oskar Klein Centre, Department of Astronomy, Stockholm University, AlbaNova, SE-106 91 Stockholm, Sweden}
\email{anjashagangopadhyay@gmail.com}

\author{G. C. Anupama}
\affiliation{Indian Institute of Astrophysics, Koramangala 2nd Block, Bangalore 560034, India}
\email{gca@iiap.res.in}



\begin{abstract}

We present near-infrared (NIR) and optical observations of the highly reddened Type IIP supernova (SN) 2023gfo  in the nearby galaxy NGC 4995 ($d = 26.4 \pm 3.2$ Mpc), which reached a high peak luminosity of $M_V = -18.6$ mag.
The SN was initially detected as a faint red event, with $B-V = 0.8$ mag at the beginning of the plateau phase.
By comparison with template, we estimate a total extinction of $A_V = 2.1$ mag.
After correcting for this extinction, we derive a peak quasi-bolometric luminosity of $(5.9 \pm 1.5)\times10^{42}$ erg s$^{-1}$, placing this event among the most luminous SNe IIP, while its plateau duration remains within the normal range.
The early-phase optical spectrum exhibits a P-Cygni profile of H$\alpha$, with a broad absorption of $V$ = $13{,}800$ km s$^{-1}$, which is among the highest observed for SNe IIP at comparable epochs.
The high luminosity and the normal plateau duration suggest that this event represents an outlier.
Applying an analytical model, we infer an unusually large progenitor radius.
This may indicate that the progenitor experienced an extreme energy injection from the core to the envelope shortly before explosion, resulting in a substantially inflated radius.
While ejecta-circumstellar matter (CSM) interaction could in principle account for the high luminosity, we find no observational evidence supporting strong interaction.

\end{abstract}


\section{Introduction} \label{sec:intro}


Massive stars \citep[$M_{\rm ZAMS} >8$--$10~\mathrm{M}_{\odot}$;][]{Heger2003} reach the end of their evolutionary stage and subsequently undergo core-collapse supernovae (CCSNe), i.e., terminal explosions of massive stars. 
Those events characterized by the presence of hydrogen lines in their optical spectra are classified as Type II SNe \citep[SNe II;][]{Minkowski1941,Filippenko1997}. 
Type IIP SNe constitute the most common subclass of CCSNe, accounting for approximately half of all observed events \citep[][]{Li2011}.
They are characterized by an extended plateau phase in their optical light curves, lasting for several months. This plateau is generally attributed to the recombination of hydrogen in the massive envelope retained by the progenitor star at the time of explosion.
Despite the growing number of observations, the connection between progenitor properties and the observable characteristics of SNe IIP remains poorly constrained and continues to be actively debated.



Observational diversity is seen in several observed properties among SNe IIP. 
Their absolute peak magnitudes in the $V$-band show a wide distribution, typically ranging from approximately $-15$ to $-18$ mag \citep[][]{Anderson2014}. This diversity extends to their temporal evolution; previous studies have established several key statistical correlations linking their photometric features. 
Furthermore, a proportional relationship between luminosity and expansion velocity has also been found \citep[][]{Hamuy2003}, emphasizing the intrinsic links between explosion energy, the amount of radioactive nickel synthesized, and the resulting photometric diversity of SNe IIP.

Recently, highly-ionized emission lines with a blue continuum were detected for some Type IIP SNe during their early phases \citep{Khazov2016,Yaron2017}. The high ionization temperature indicates that the surrounding gas is ionized by the strong radiation field emanating from the central SN. 
This `flash feature' infers that a progenitor is surrounded by a dense and compact CSM located at $\sim 10^{14}$--$10^{15}$ cm, frequently referred to as the `confined CSM'. 
Given the typical ejecta velocity, the ejecta quickly sweeps up the confined CSM. In such a situation, the interaction power causes a faster rise and bumpy structure in the early-phase light curve \citep{Moriya2024}. 

One possible scenario of confined CSM is enhanced mass loss driven by increased stellar activity in the final stages of stellar evolution.
However, the observational consequences of such late-stage energy deposition are not yet fully understood.
In some cases, it could lead to a substantial radius expansion rather than the formation of the confined CSM, which is predicted to result in a peculiar SN IIP with a high luminosity and long plateau as compared to canonical SNe IIP \citep{Oucih2019}. \citet{Ouchi2021} showed that the observational properties of the peculiar SN IIP 2009kf, which has these properties \citep{Botticella2010}, could be explained within this scenario and derived the progenitor radius significantly larger than those of typical Type IIP SNe. 

In this paper, we present optical and NIR observations of SN 2023gfo, a high-luminosity Type IIP SN. 
Section 2 details the telescopes and instruments used for the observations.
Section 3 provides a detailed description of the extinction correction applied to the data. Then, we present the results, including the light curves, spectral data, and the bolometric light curve derived from our observations, comparing them with those of Type IIP SNe previously studied. In section 4, we discuss the physical parameters of SN 2023gfo, where we find that the progenitor radius is larger than those of typical Type IIP SNe; a similar situation is found in SN 2009kf, and we suggest that the peculiar nature of SN 2023gfo could be explained by an inflated progenitor star caused by the final stellar activity. 
Section 5 summarizes our conclusions.

\section{OBSERVATIONS AND DATA REDUCTION} \label{sec:style}

\begin{figure}
    \centering
\includegraphics[width=0.7\textwidth]{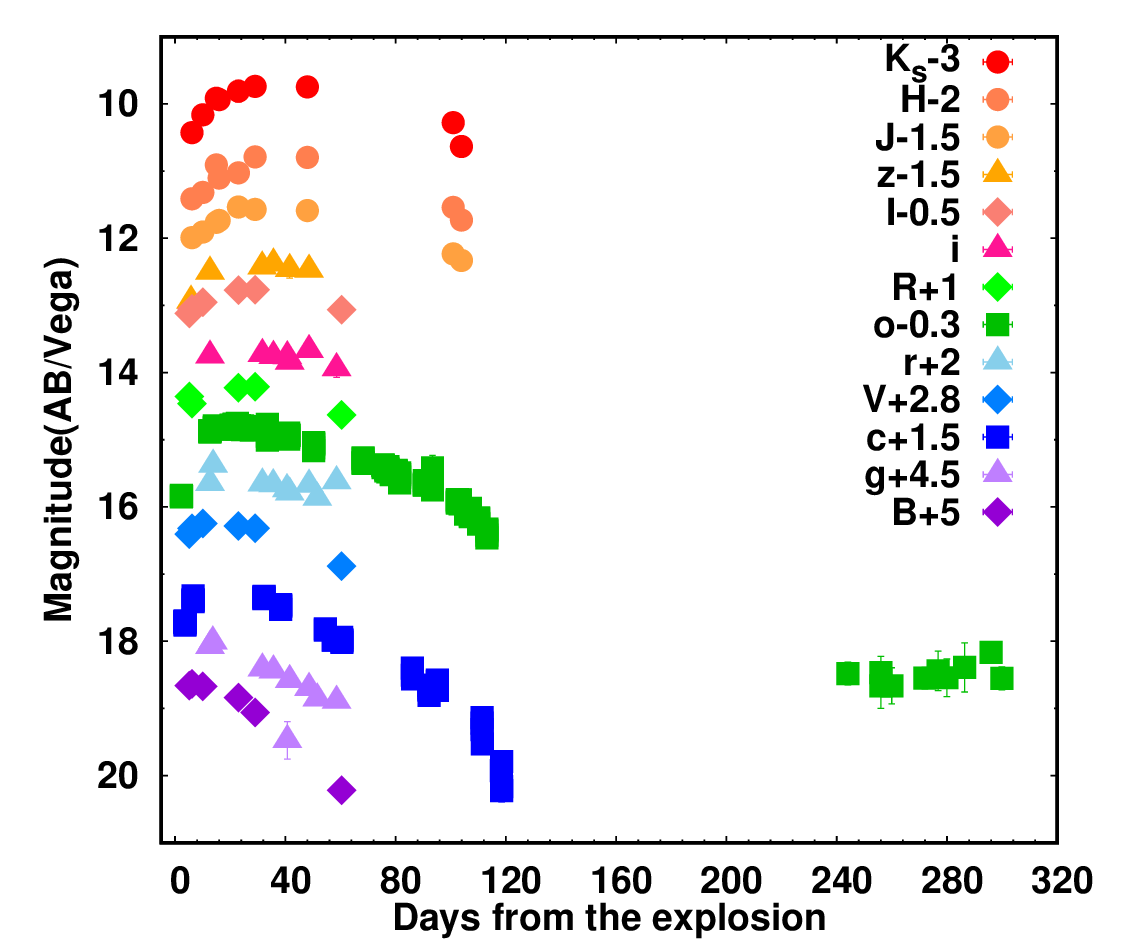}
    \caption{The optical and NIR light curve of SN 2023gfo. 
    Photometric data are color-coded and plotted with various symbols: 
  ATLAS $c$ and $o$ bands (squares), 
  GIT $g, r, i,$ and $z$ bands (triangles), 
  HOWPol $B, V, R,$ and $I$ bands (diamonds), and 
  kSIRIUS $J, H,$ and $K_{\rm s}$ bands (circles). 
The $B$, $V$, $R$, $I$, $J$, $H$, and $K_s$-band magnitudes are in the Vega system, and $c$, $o$, $g$, $r$, $i$, and $z$-band ones in the AB system. 
 Each magnitude is artificially shifted to distinguish the light curves.}
\end{figure}

\begin{figure}
\centering
   \includegraphics[width=0.7\textwidth]{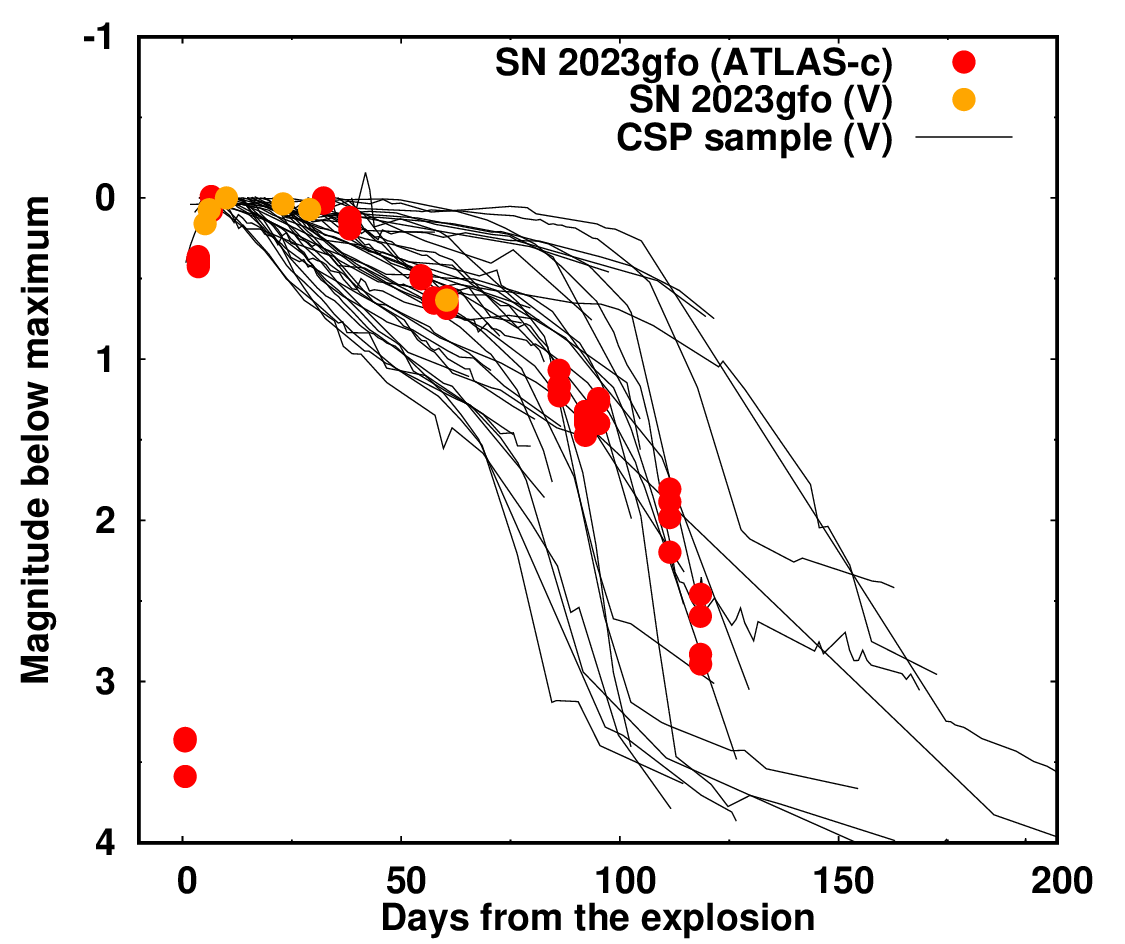}
   \caption{The ATLAS $c$-band and HOWPol $V$-band light curve of SN 2023gfo compared with the $V$-band ones of Type IIP SNe in the CSP sample \citep[][]{Anderson2014}, normalized to the peak magnitude (set to zero). }
   \end{figure}


 \begin{figure}
 \centering
    \includegraphics[width=0.8\textwidth]{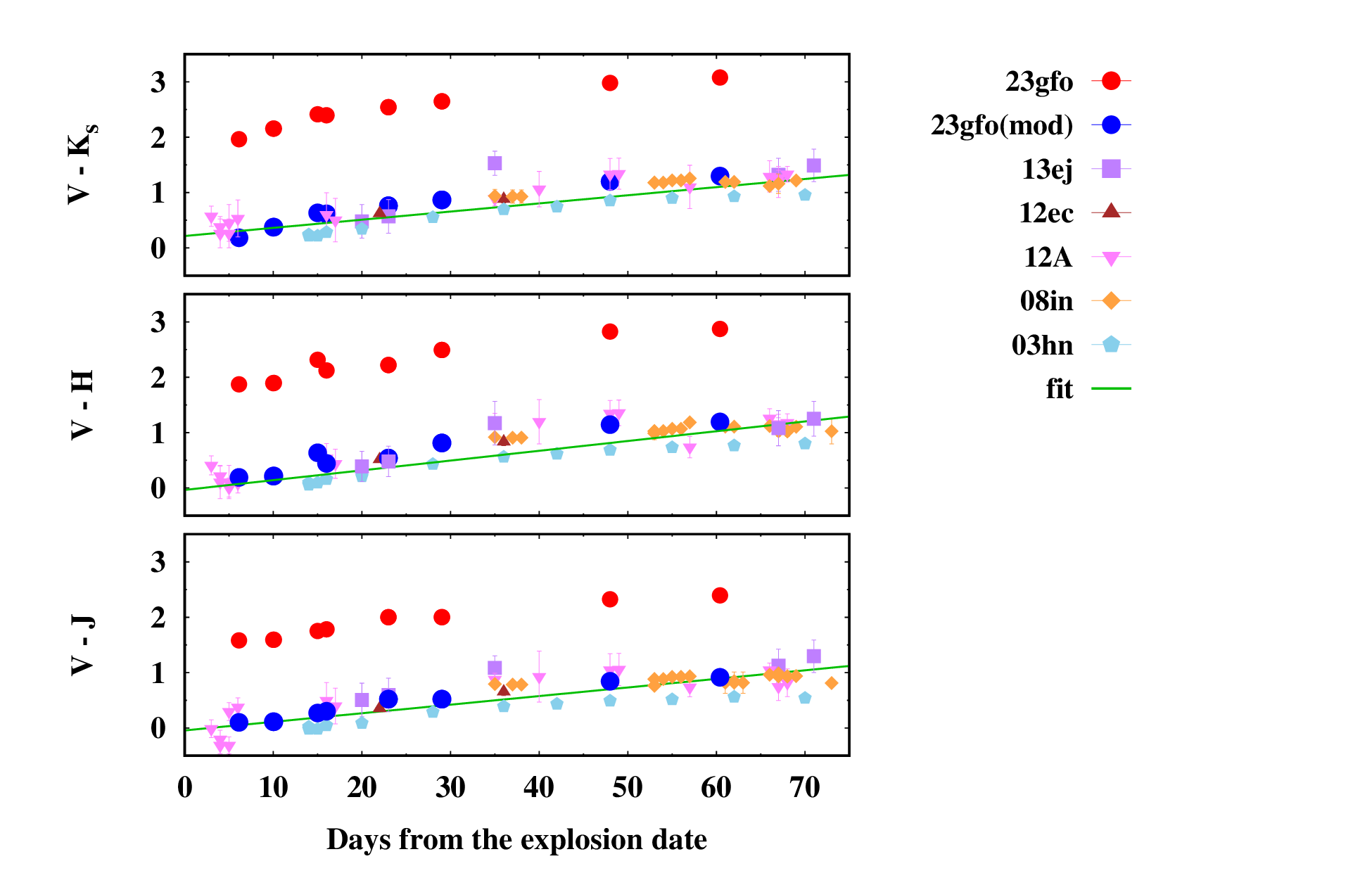}
    \caption{The $V$-$J$$H$$K_{s}$ color evolution of SN 2023gfo compared
    with those of other SNe 2003hn \citep[][]{Kevin2009}, 2008in \citep[][]{Roy2011}, 
    2012A \citep[][]{Tomasella2013}, 2012ec \citep[][]{Barbarino2015}, and 2013ej \citep[][]{Huang2015}.
   Blue points represent SN 2023gfo before the extinction correction, while red points show the extinction-corrected SN 2023gfo with $A_V=2.1$ mag. The green-color solid line is represents a linear function fitting to combined data of    comparative SNe.}
    
 \end{figure}

 \begin{figure*}[htbp]
  \centering
  \begin{tabular}{cc}
   \includegraphics[width=0.5\textwidth]{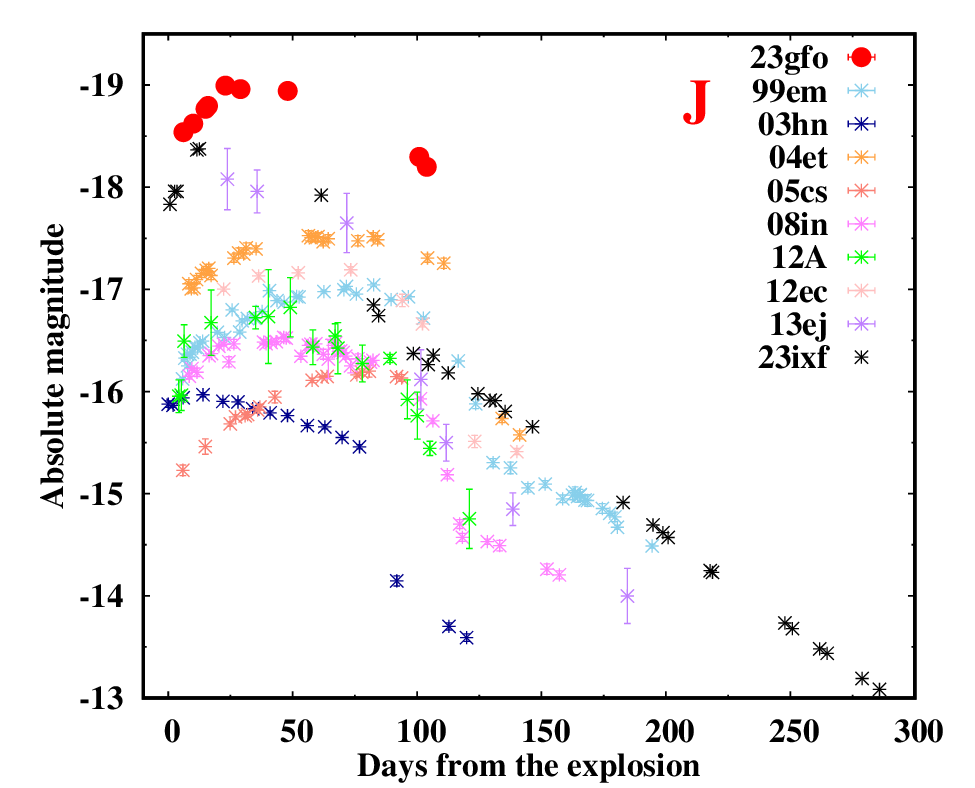} & 
    \includegraphics[width=0.5\textwidth]{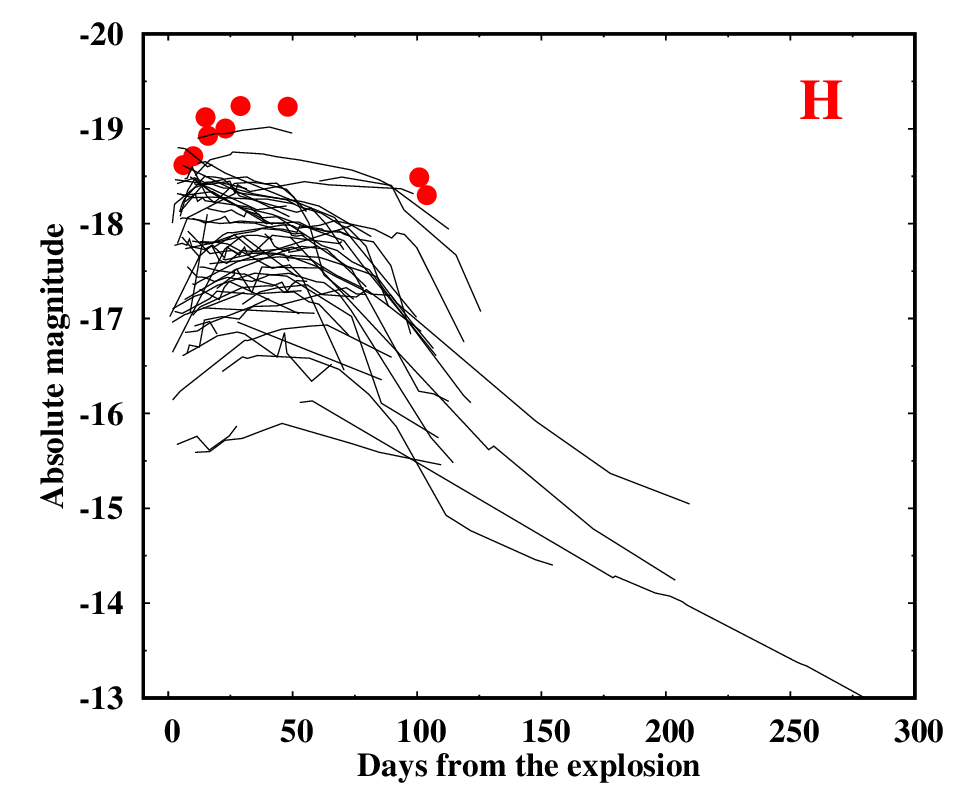} \\     \includegraphics[width=0.5\textwidth]{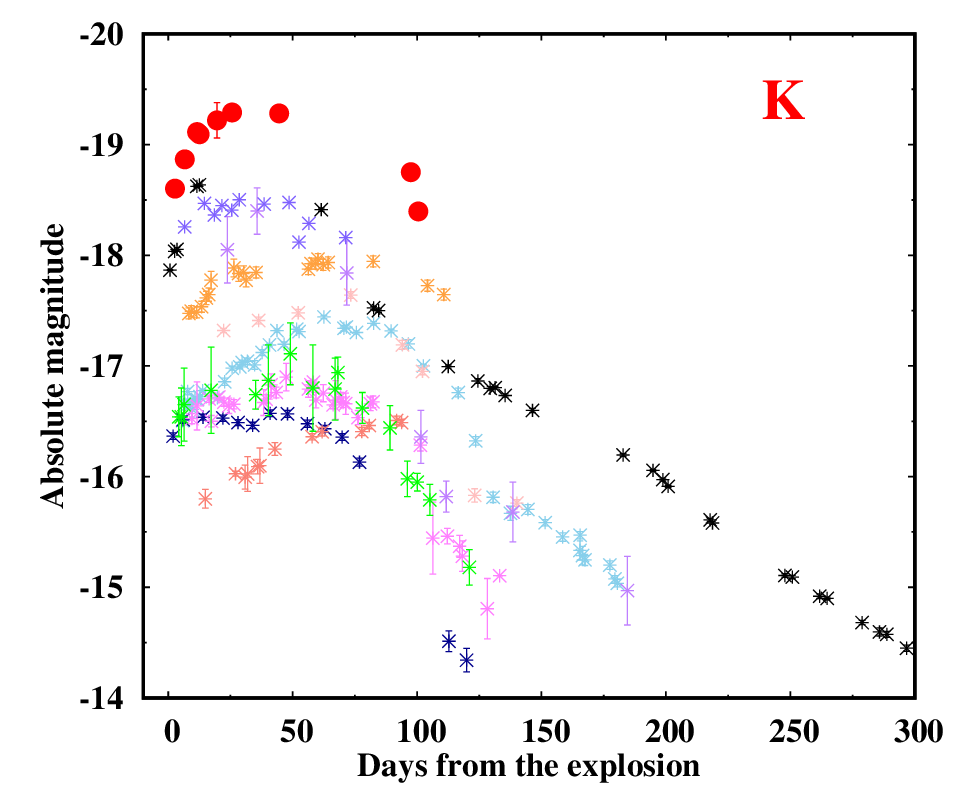} & \includegraphics[width=0.5\textwidth]{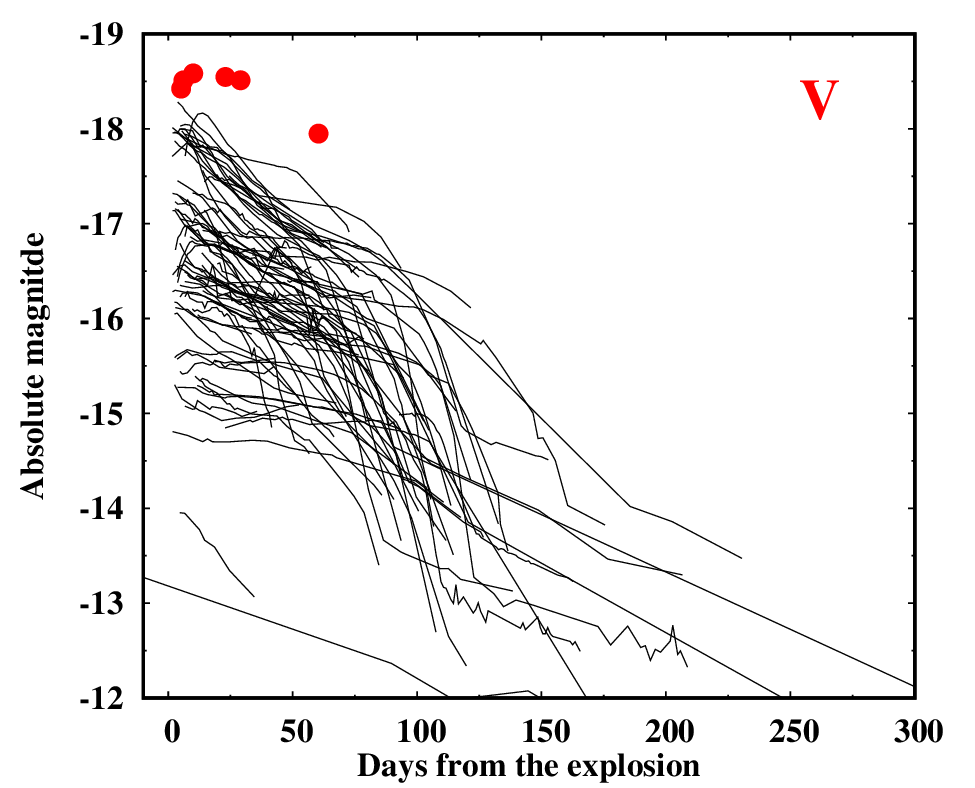}
    \end{tabular}
  \caption{The $J$ (left top), $H$ (right top), $K_{s}$ (left bottom), and $V$ band (right bottom) absolute magnitude light curves of SN 2023gfo. The $J$ and $K_{s}$ band light curves are compared with those of SNe 1999em \citep[][]{Elmhamdi2003}, 2003hn \citep[][]{Kevin2009}, 2004et \citep[][]{Maguire2010}, 2005cs \citep[][]{Pastorello2009}, 2008in \citep[][]{Roy2011}, 2012A \citep[][]{Tomasella2013}, 2012ec \citep[][]{Barbarino2015}, 2013ej \citep[][]{Huang2015}, and 2023ixf \citep[][]{Yamanaka2023,Singh2024}. 
  For the $V$ and $H$-band, the samples from \citep[][]{Martinez2022} are shown for comparison. The red points in each graph denote SN 2023gfo. The absolute magnitudes of SN 2023gfo were corrected using the extinction values obtained in this study ($A_V= 2.1$ mag, $A_J= 0.51$ mag, $A_H= 0.3$ mag, and $A_{K_s}= 0.17$ mag), while those of other objects were corrected using the extinction values and distances from their respective literatures.}
  \label{fig:2x2_figures}
 \end{figure*}

\begin{figure}[htbp]
\centering
\noindent
\begin{minipage}{0.6\textwidth}
  \centering
    \includegraphics[width=\linewidth]{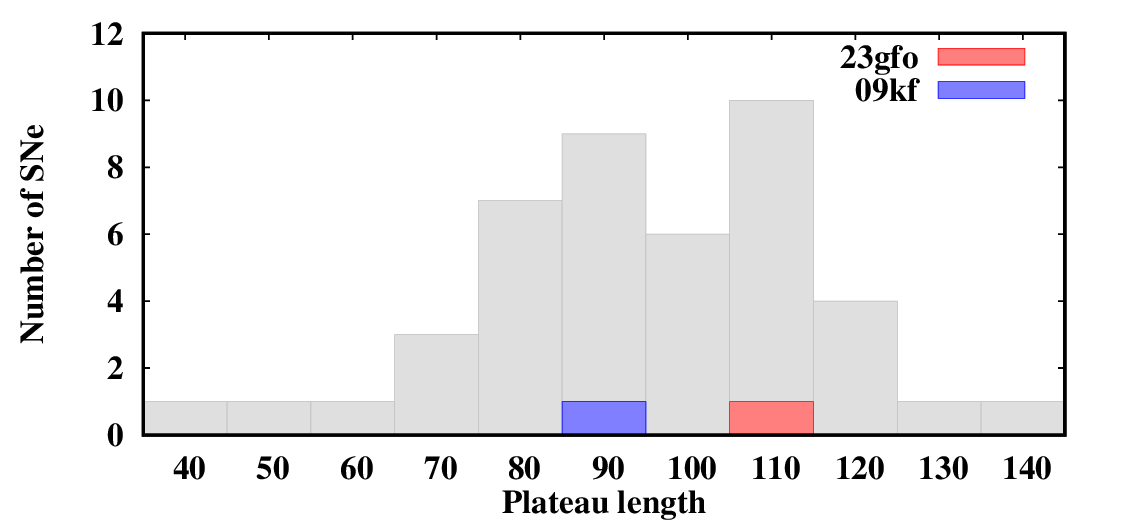}\\[1ex]
  \includegraphics[width=\linewidth]{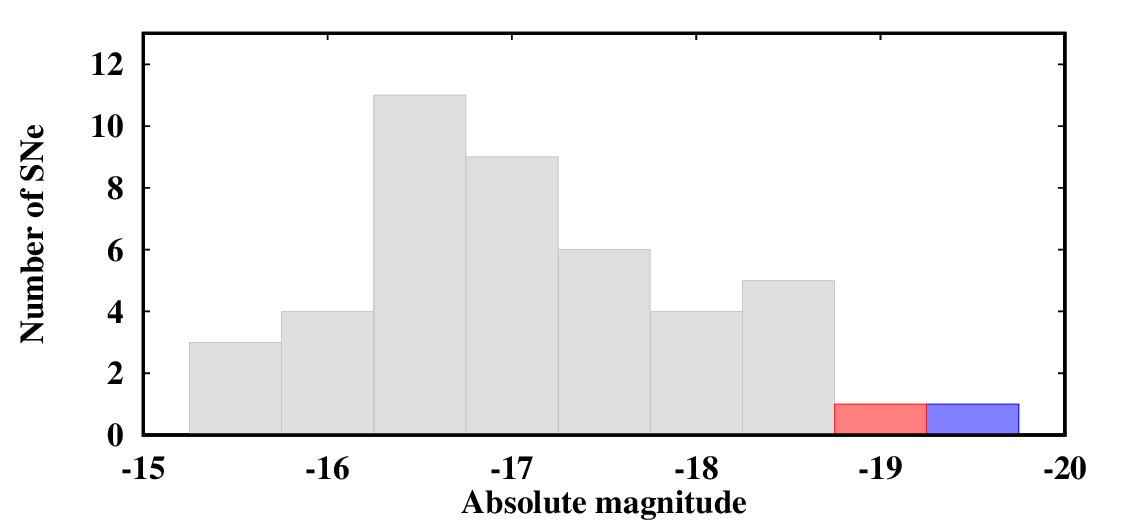}
  \caption{Top: Distribution of the plateau length, as defined by \citet{Anderson2014}, for the CSP Type II SN sample. The positions of SN 2023gfo and SN 2009kf are indicated in red and blue, respectively.
Bottom: Distribution of the peak absolute magnitudes in the $V$ band for the same sample, with SN 2023gfo and SN 2009kf highlighted.}
  \label{}
\end{minipage}
\end{figure}




\begin{figure}
\centering
    \includegraphics[width=0.6\textwidth]{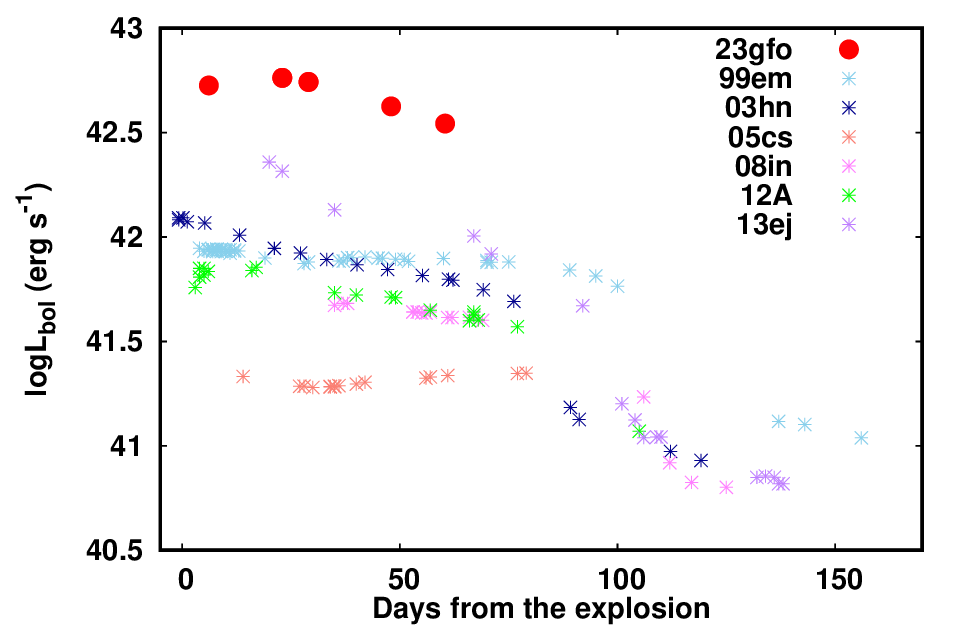}
    \caption{The bolometric light curve of SN 2023gfo compared with those of 
    SNe 1999em \citep[][]{Elmhamdi2003}, 2003hn \citep[][]{Kevin2009}, 2005cs \citep[][]{Pastorello2009}, 2008in \citep[][]{Roy2011}, 2012A \citep[][]{Tomasella2013}, and 2013ej \citep[][]{Huang2015}. The red point represents SN 2023gfo. The horizontal axis shows days since explosion, and the vertical axis shows the logarithm of the bolometric luminosity.}
 \end{figure}

\begin{figure*}[htbp]
\centering
  \begin{tabular}{cc}
     \includegraphics[width=0.5\textwidth]{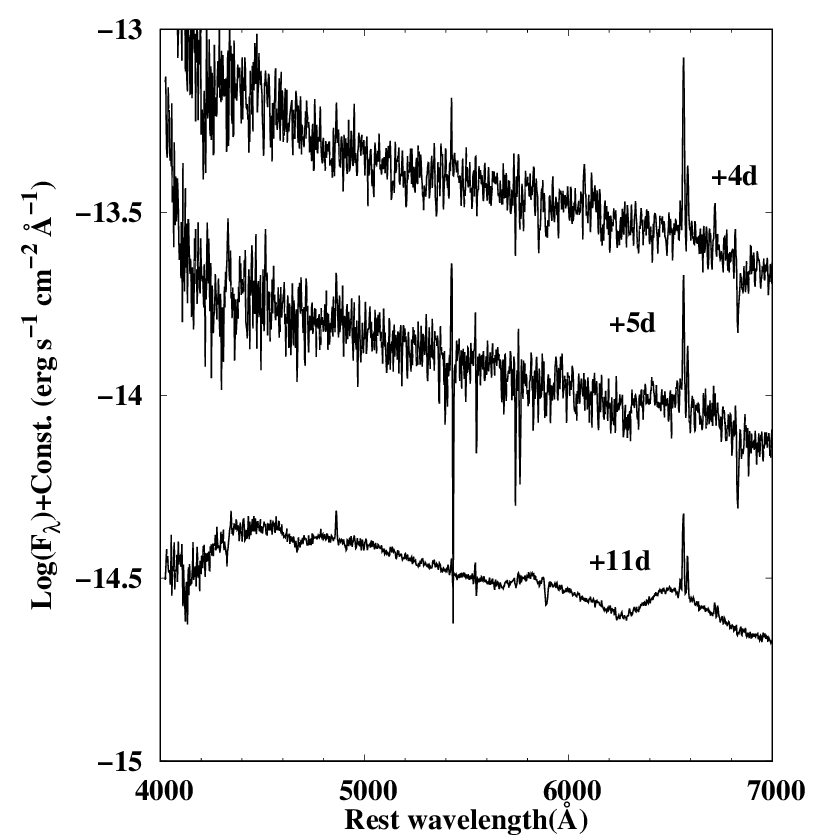}
    \includegraphics[width=0.5\textwidth]{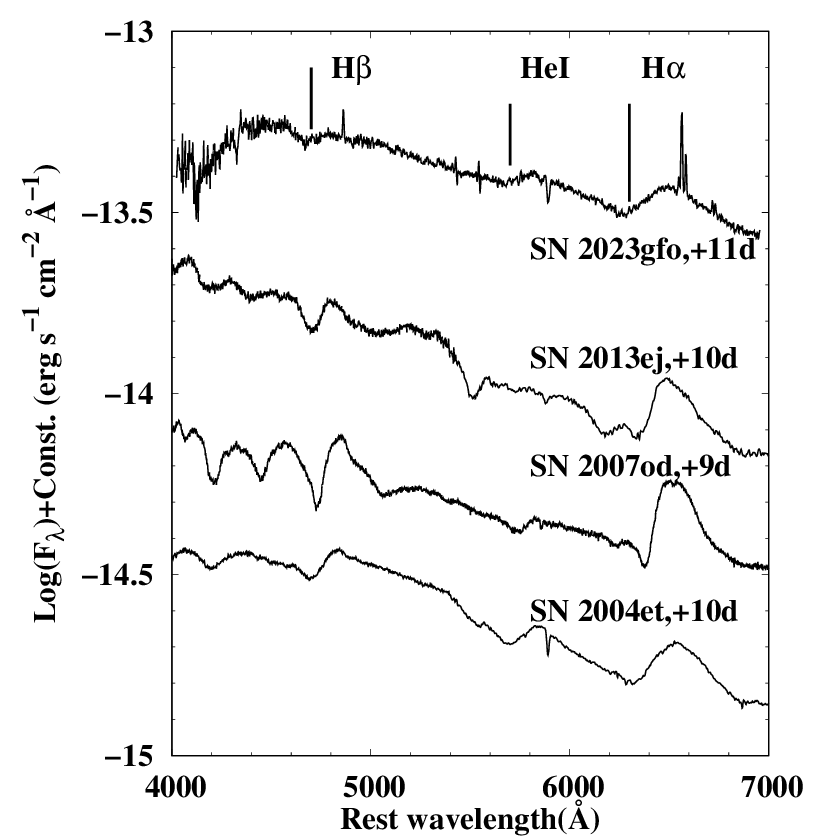}
  \end{tabular}{}
    \caption{
    (Left) Optical spectra of SN 2023gfo obtained at $t = 4$ d, 5 d, and 11 d, displayed from top to bottom, respectively.
The flux has been corrected for the total line-of-sight extinction, dominated by the host-galaxy component, using $A_V = 2.1$~mag.
The spectra are shown in the rest frame of the host galaxy, determined from the narrow host-galaxy emission lines.
    (Right) Optical spectrum of SN 2023gfo at t = 11~d, compared with spectra of Type IIP SNe 2007od \citep[][]{Inserra2011}, 2004et \citep[][]{Sahu2006}, and 2013ej \citep[][]{Valenti2014} as similar phases.}
    \label{fig:2x2_fig1}
\end{figure*}

SN 2023gfo was discovered at 16.2 magnitude by the Asteroid Terrestrial Impact Last Alert System (ATLAS; \citealt{Tonry2018}; \citealt{Smith2020}) on 20 April 2023 (UT) (MJD 60054.2) \citep{Tonry2023}. 
Prior to discovery, this object was detected at 19.2 mag on MJD 60051.2, while the last non-detection was reported at MJD 60050.0 with a limiting magnitude of 19.5 mag \citep{Tonry2023}.
These observations indicate a rapid rise shortly after the explosion.
Throughout this paper, we adopt an explosion epoch of MJD 60050.6 (t= 0).
Spectroscopic observations subsequently classified SN 2023gfo as a Type IIP SN, and revealed significant reddening along the line of sight \citep{Fulton2023}.
\cite{Yang2023} independently classified SN 2023gfo and adopted a host-galaxy redshift of $z = 0.00594$, corresponding to a recession velocity of approximately 1780 km s$^{-1}$.
The redshift-independent distance of the host galaxy NGC 4995, primarily obtained using the Tully-Fisher (TF) method from the NASA/IPAC Extragalactic Database (NED), is $26.4 \pm 3.2$ Mpc. The corresponding distance modulus is $32.06 \pm 0.42$ mag \citep{Tully2016,Springob2009,Springob2009a,Karachentsev2013,Sorce2014,Tully2013,Sorce2014a}. 
Our adopted TF distance is consistent with the Hubble-flow (HF) distance 
provided by the NASA/IPAC Extragalactic Database (NED). According to NED, the HF distance corrected for the Virgo Cluster, the Great Attractor (GA), and the Shapley Supercluster is $m-M = 32.35 \pm 0.15$ mag. The agreement between these two independent distance indicators reinforces the reliability of our distance scale. We adopted these values throughout this paper. 
 
Optical photometric observations in the $BVRI$-bands were conducted using HOWPol \citep{Kawabata2008} mounted on the 1.5-m Kanata Telescope located at Higashi-Hiroshima Observatory. 
Additional optical photometry in the $g$,$r$,$i$,$z$ bands was obtained with the GROWTH-India Telescope \citep[GIT; ][]{Kumar2023} at the Indian Astronomical Observatory (IAO) in Hanle.
We monitored SN 2023gfo in the NIR-band using the kSIRIUS, which is the simultaneous NIR $JHK_{\rm s}$ band imager \citep[][]{Nagayama&Nakaya2024} attached to the Cassegrain focus of the Kagoshima 1.0-m Telescope at the Iriki Observatory.

All imaging data were reduced following standard procedures using the Image Reduction and Analysis Facility (IRAF) \citep[][]{Tody1986}. The photometric magnitudes were obtained through point-spread-function photometry using standard IRAF tasks such as DAOPHOT \citep[][]{Stetson1987}. For the optical observations, the photometric calibration was performed using secondary stars from the Pan-STARRS catalog \citep[][]{Flewelling2020}. For the NIR observations, the calibration was performed using the two-micrometer all-sky survey catalog \citep[2MASS;][]{Skrutskie2006}.

Optical spectroscopic observations were conducted over three nights using KOOLS-IFU \citep{Matsubayashi2025} on the 3.8-m Seimei Telescope at the Okayama Observatory \citep{Kurita2020}. 
The data reduction was performed following standard procedures using IRAF \citep{Tody1986,Tody1993}. Standard corrections for atmospheric extinction and telluric absorption were applied.
IRAF is distributed by the National Optical Astronomy Observatory, which is operated by the Association of Universities for Research in Astronomy (AURA) under a cooperative agreement with the National Science Foundation.
\footnote{IRAF (Image Reduction and Analysis Facility) was originally distributed by the National Optical Astronomy Observatory (NOAO), which was operated by the Association of Universities for Research in Astronomy (AURA) under a cooperative agreement with the National Science Foundation. IRAF is currently maintained by NSF NOIRLab. See \url{https://iraf.noirlab.edu}.
 The wavelength scale was corrected for the host-galaxy recession velocity corresponding to a redshift of $z = 0.00594$.
}

\section{RESULTS}
\subsection{Optical and infrared light curves}
Photometric monitoring of SN 2023gfo began at t= 6 d (see Figure 1). 
The $V$-band light curve showed a slight rise from $13.6 \text{ mag}$ at t= 5.2 d to a peak magnitude of $13.5 \text{ mag}$ at t= 10.0 d, followed by a subsequent decline (see Figure 1).
In the NIR, the $J$-band light curve reached its maximum at approximately at t= 23 d. The maximum date was reached later at longer wavelengths.

Figure 2 compares the $c$-band light curve of SN 2023gfo with the $V$-band light curves of the CSP sample \citep{Anderson2014}, each normalized by its peak magnitude. It is within the distribution of the light curves of the CSP sample.
The comparison between the ATLAS $c$-band and the literature $V$-band samples is justified because the $c$-band wavelength range encompasses the standard $V$-band \citep[][]{Tonry2018}, and 
the $c$-band light curve distribution of SNe IIP is very similar
to that of the $V$-band \citep{Ertini2026}.
Following the methodology of \cite{Anderson2014}, we derived the plateau duration using the $c$-band light curve, finding a length of $108 \pm 8$ days. 
 Therefore, in this study, we adopt a plateau duration of $\sim108$ days.

\subsection{Extinction}


SN 2023gfo exhibits significant reddening, requiring a correction for extinction. As an initial estimate, we measured
the equivalent width (EW) of the Na~I~D absorption lines at the host-galaxy redshift ($z=0.00594$), which yielded a value of $\sim2.4$~\AA, indicative of substantial extinction \citep{Poznanski2012}. However, at such large EW values, the empirical relation proposed by \citet{Poznanski2012} becomes unreliable.


We therefore applied another method to determine the extinction, primarily through a comparison of the observed color evolution with those of template Type IIP SNe.
The optical spectrum obtained at t = 11 d shows typical absorption features of a normal Type IIP SN (see \S 3.6). The characteristic light-curve evolution is also consistent with typical SNe IIP (see Figure 2).
These properties suggest that the plateau phase is shaped by hydrogen recombination. In this scenario, the photospheric temperatures of SNe IIP are expected to be similar, leading to relatively uniform intrinsic colors \citep[][]{Martinez2022}. Based on this assumption, we applied methods for correcting extinction using color evolution (Figure 3).

We first estimated the $V$-band extinction using the $B-V$ color at 30 days after the explosion.
Following the prescription of \citet{Jaeger2018}, this method yielded $E(B-V) = 0.52 \pm 0.19$ mag (see also \citep[][]{Yang2023}), corresponding to $A_V = 1.6 \pm 0.6$ mag assuming $R_V = 3.1$ \citep[][]{Cardelli1989}, although with a relatively large uncertainty.


To obtain a more robust estimate, we constructed color curves by combining $V$ and $JHK_s$-band photometry. For the intrinsic $V$-$JHK_s$ colors, we adopted template color derived from linear fits to merged data of SNe 2003hn \citep[][]{Kevin2009}, 2008in \citep[][]{Roy2011}, 2012ec \citep[][]{Barbarino2015}, 2012A \citep[][]{Tomasella2013}, and 2013ej \citep[][]{Huang2015}, using data until t=70 d. Using these templates, we derived color excesses for SN 2023gfo of $E(V-J)=1.6 \pm 0.1$ mag, $E(V-H)=1.9 \pm 0.1$ mag, and $E(V-K_s)=1.9 \pm 0.1$ mag.
Adopting a Galactic extinction law ($R_V = 3.1$), these values correspond to an extinction of $A_V = 2.1 \pm 0.1$ mag.
It is important to note that the NIR magnitudes are significantly less sensitive to the choice of $R_V$ compared to the optical $V$-band. For instance, even with $A_V = 2.1$ mag, the extinction in the $K_s$-band is as small as $A_{K_s}=0.17$ mag. This low sensitivity to $R_V$ uncertainties ensures a more robust derivation of the intrinsic luminosity.
Given the smaller uncertainty associated with the $V$-$JHK_s$ estimate, we adopt $A_V = 2.1$ mag to correct all observational data presented in this study.
As independent support for this extinction value, the $A_V$-corrected spectra exhibit an $H\alpha/H\beta$ flux ratio that is consistent with the intrinsic Balmer decrement expected for Case B recombination \citep{Osterbrock2006}. This agreement confirms the validity of our adopted $A_V = 2.1$ mag.

\subsection{Absolute magnitude}
We derived the absolute magnitudes of SN 2023gfo after correcting for extinction and distance modulus. For comparison, we compiled photometric data of SNe 1999em \citep[][]{Elmhamdi2003}, 2003hn \citep[][]{Kevin2009}, 2004et \citep[][]{Maguire2010}, 2005cs \citep[][]{Pastorello2009}, 2008in \citep[][]{Roy2011}, 2012A \citep[][]{Tomasella2013}, 2012ec \citep[][]{Barbarino2015}, 2013ej \citep[][]{Huang2015}, and 2023ixf \citep[][]{Yamanaka2023,Singh2024}. The resulting absolute-magnitude light curves are shown in Figure 4. 
In the NIR bands, SN 2023gfo is significantly brighter than the comparison Type IIP SNe.
In particular, its peak absolute magnitude reaches $M_{K_s} = -19.5\pm 0.4$ mag. 
Consistently, SN 2023gfo is also brighter in the $J$ and $H$ bands than all comparison objects.
In the optical, we further compared the $V$-band absolute magnitude of SN 2023gfo with the sample presented by \citet{Anderson2014}.
SN 2023gfo reaches a peak absolute magnitude of $M_V = -18.6 \pm 0.4$ mag, placing it among the most luminous Type IIP SNe observed to date.
For reference, even the most luminous events in the \citet{Anderson2014} sample reach peak magnitudes of $M_V = -18.3~\pm~0.4$ mag. These comparisons demonstrate that SN 2023gfo is an exceptionally luminous Type IIP SN.

We examined the plateau duration and the peak absolute magnitude of SN 2023gfo relative to the distribution of $V$-band light curves in the CSP sample (see Figure 5). For comparison, the values for SN 2009kf are also shown.
We find that the plateau duration of SN 2023gfo and SN 2009kf is close to the mode of the CSP distribution, indicating plateau durations that are typical of Type IIP SNe. In contrast, their absolute magnitudes are located at the extreme luminous end of the distribution.
In summary, while SN 2023gfo exhibits a plateau duration and light-curve morphology consistent with those of normal Type IIP SNe, its peak luminosity is exceptionally high compared to the bulk of the population.

\subsection{Bolometric light curve}
Using optical and NIR photometry, we constructed the bolometric light curve of SN 2023gfo based on the extinction-corrected magnitudes. 
The observed magnitudes were converted to flux densities at the effective wavelengths of each filter \citep{Fukugita1996}, and the resulting spectral energy distribution was integrated over wavelength to obtain the bolometric luminosity.
In this study, the bolometric luminosity is defined as a pseudo-bolometric luminosity derived from the $BVRIJHK_{s}$ bands.
For consistency, we derived pseudo-bolometric light curves for other Type IIP SNe using the same filter set and methodology.
Figure 6 compares the $BVRIJHK_{s}$ pseudo-bolometric light curve of SN 2023gfo with those of other Type IIP SNe.
SN 2023gfo reaches a maximum bolometric luminosity of $(5.9 \pm 1.5) \times 10^{42}$ erg s$^{-1}$ at $t = 23.6$ d.
This peak luminosity is significantly higher than those of the comparison objects at similar phases.

\subsection{Optical Spectroscopy}
Optical spectroscopic observations were obtained at $t = 4$ d, $5$ d, and $11$ d after the estimated explosion date (Figure 7).
In the following, we focus on the spectrum at $t = 11$ d, 
which allows detailed analysis of spectral features thanks to its high signal-to-ratio. 
The $t = 11$ d spectrum of SN 2023gfo is compared with those of SNe 2007od \citep[][]{Inserra2011}, 2004et \citep[][]{Sahu2006}, and 2013ej \citep[][]{Valenti2014} at similar phases.
The spectrum exhibits a prominent P-Cygni profile in H$\alpha$, with an expansion velocity of approximately 13,800 km s$^{-1}$. Broad absorption features of H$\beta$ and He~{\sc I} $\lambda5876$ are also clearly detected.
The inferred expansion velocity is among the highest reported for Type IIP SNe at comparable epochs.

The spectra shows narrow emission lines of $\text{H}\alpha$, [\text{N}~\textsc{ii}], and [\text{S}~\textsc{ii}], which originates from the underlying $\text{H}~\textsc{ii}$ region within the host galaxy. We measured the ratios of these emission line strengths. Based on the method described in \citet{PP04}, we estimated the metallicity at the SN site to be 8.7, and found it to be consistent with the average value for Type II SNe \citep{Anderson2010}.



\section{DISCUSSION} \label{sec:pubcharge}

Section 3 highlighted the key properties of SN 2023gfo, including the plateau duration, expansion velocity, and bolometric luminosity.
SN 2023gfo is characterized by a high luminosity ($M_{\text{V}}=-18.6~\text{mag}$) and a high expansion velocity ($v_{\text{H}\alpha}=13,800~\text{km}~\text{s}^{-1}$), despite its normal plateau length ($t_{\text{p}}=108~\text{d}$). 
This combination places SN 2023gfo as an outlier relative to the distribution of Type IIP SNe.



Although SN 2023gfo lies at the luminous end of the Type IIP population, \citet{Oucih2019} demonstrated through radiation-hydrodynamic calculations that the analytic scaling relations of \citet{Popov1993} and \citet{Kasen&Woosley2009} successfully reproduce the plateau luminosities and durations of Type IIP SN models with expanded envelopes. This result supports the applicability of these scaling relations even in the high-luminosity regime represented by SN 2023gfo.
We therefore estimated the physical properties of SN 2023gfo using the analytic scaling relations of \citet{Popov1993} and \citet{Kasen&Woosley2009}. As a primary reference object, we adopted SN 2009kf, whose observational properties closely resemble those of SN 2023gfo. SN 2009kf exhibited an unusually high luminosity and a large expansion velocity while maintaining a plateau duration comparable to those of normal Type IIP SNe \citep{Botticella2010}. Furthermore, \citet{Ouchi2021} derived its physical properties through radiation-hydrodynamic modeling and reported a progenitor radius exceeding $6000~R_{\odot}$.


Using SN 2009kf as the primary reference object, 
we estimated the explosion and progenitor properties of SN 2023gfo using the analytic scaling relations of \citet{Popov1993} and \citet{Kasen&Woosley2009}:

\begin{equation}
L_{\rm SN} \propto E_{\rm k}^{5/6} M_{\rm ej}^{-1/2} R_0^{2/3} \kappa^{-1/3} T_{\rm I}^{4/3} \ ,
\end{equation}
\begin{equation}
t_{\rm SN} \propto E_{\rm k}^{-1/6} M_{\rm ej}^{1/2} R_0^{1/6} \kappa^{1/6} T_{\rm I}^{-2/3} \ ,
\end{equation}
\begin{equation}
v_{\text{SN}} \approx \left(\frac{2E_{\text{k}}}{M_{\text{ej}}}\right)^{1/2} \
\end{equation}
where $L_{\rm SN}$ and $t_{\rm SN}$ are the luminosity and duration of the plateau phase. $v_{\text{SN}}$ denotes the ejecta velocity. $E_{\rm k}$, $M_{\rm ej}$, and $R_0$ denote the explosion energy, ejecta mass, and pre-SN progenitor radius, respectively. $\kappa$ and $T_{\rm I}$ are the opacity and ionization temperature, respectively, which can be taken to be universal among different objects. 


The H$\alpha$ absorption minimum of SN 2023gfo was measured to be
13,800 km s$^{-1}$ at $t=11$ d.
Since spectra around the middle of the plateau phase were not available,
we estimated the expansion velocity at $t=60$ d by fitting the velocity
evolution of SN 2017eaw using an exponential function and scaling it to match the observed velocity of
SN 2023gfo.
The resulting velocity at $t=60$ d was approximately
7000 km s$^{-1}$ and was adopted for the scaling analysis.

For the comparison objects, velocities measured at similar phases were adopted. The bolometric luminosity was $(3.5 \pm 1.5) \times 10^{42}$ erg s$^{-1}$ at $t = 60$ d for SN 2023gfo. The luminosities at similar phases were used for comparison SNe, ensuring a consistent comparison. 

\begin{table}[ht]
\centering
\caption{Observed properties adopted for the scaling analysis}
\label{tab:obs}
\begin{tabular}{lccc}
\hline
SN &
$v_{\rm 60}$ &
$t_{\rm p}$ &
$L_{\rm bol}(60)$ \\
 &
(km s$^{-1}$) &
(d) &
(erg s$^{-1}$) \\
\hline
SN 2009kf & 9000 & $\sim90$ & $6.3\times10^{42}$ \\
SN 2023gfo & 7000 & 108 & $3.5\times10^{42}$ \\
\hline
\end{tabular}
\end{table}

\begin{table}[ht]
\centering
\caption{Physical properties of SN 2009kf and the scaled values for SN 2023gfo}
\label{tab:phys}
\begin{tabular}{lccc}
\hline
SN &
$E_{\rm k}$ &
$M_{\rm ej}$ &
$R_0$ \\
 &
(foe) &
($M_\odot$) &
($R_\odot$) \\
\hline
SN 2009kf & 2.8 & 11 & 6000 \\
SN 2023gfo & 3.0 & 19.3 & 3500 \\
\hline
\end{tabular}
\end{table}

\begin{table*}[ht]
\centering
\caption{Physical parameters adopted for the reference SNe and the corresponding parameters derived for SN~2023gfo.}
\label{tab:scaling}

\begin{tabular*}{\textwidth}{@{\extracolsep{\fill}}lccccccc}
\hline
&
\multicolumn{3}{c}{Reference SN}
&
\multicolumn{3}{c}{Derived for SN 2023gfo}
\\

Reference SN
&
$E_{\rm k}$
&
$M_{\rm ej}$
&
$R_0$
&
Ref.
&
$E_{\rm k}$
&
$M_{\rm ej}$
&
$R_0$
\\

&
(foe)
&
($M_\odot$)
&
($R_\odot$)
&
&
(foe)
&
($M_\odot$)
&
($R_\odot$)
\\

\hline
SN 2009kf
& 2.8
& 11
& 6000
& (1)
& 3.0
& 19
& 3500
\\

SN 1999em
& 1.25
& 19
& 800
& (2)
& 1.8
& 15
& 3900
\\

SN 2012A
& 0.48
& 12.5
& 259
& (3)
& 0.6
& 11
& 3600
\\

SN 2012aw
& 1.5
& 20
& 430
& (4)
& 2.0
& 13
& 3000
\\

SN 2017eaw
& 2.7
& 15
& 300
& (5)
& 2.2
& 7
& 2600
\\

\hline

Mean
& ---
& ---
& ---
& ---
& 1.92
& 13.0
& 3320
\\

Std. dev.
& ---
& ---
& ---
& ---
& 0.88
& 4.7
& 508
\\
\hline
\end{tabular*}
\vspace{2mm}

\textit{References.}
(1) \citep[][]{Ouchi2021};
(2) \citep[][]{Bersten2011};
(3) \citep[][]{Tomasella2013};
(4) \citep[][]{Dallora2014};
(5) \citep[][]{VanDyk2019}.
\end{table*}

The inferred progenitor radius is substantially larger than those typically expected for generally red supergiant progenitors of Type IIP SNe. However, a similarly large progenitor radius was also derived for SN 2009kf by \citet{Ouchi2021}.Moreover, the explosion energies inferred for SN 2023gfo and SN 2009kf are comparable, while the ejecta mass estimated for SN 2023gfo is somewhat larger. Despite this difference, both events are characterized by unusually large progenitor radius.
To evaluate the sensitivity of the inferred parameters to the choice of reference object, we additionally applied the same scaling relations using several well-studied Type IIP SNe whose physical properties have been independently constrained through progenitor detections and/or hydrodynamic and radiative-transfer modeling. The comparison sample consisted of SN 1999em \citep[][]{Elmhamdi2003,Bersten2011}, SN 2012A \citep[][]{Tomasella2013}, SN 2012aw \citep[][]{Bose2013, Dallora2014}, and SN 2017eaw \citep[][]{VanDyk2019}.
Applying the scaling relations to these objects yielded average physical parameters of $E_{\rm k}=(1.9\pm0.9)\times10^{51}$ erg, $M_{\rm ej}=13\pm4.5~M_{\odot}$, and $R_{0}=3300\pm500~R_{\odot}$ for SN 2023gfo.
The remaining dispersion likely reflects both the diversity of the reference SNe and the simplifying assumptions inherent in the analytic scaling relations, which describe the explosion using a limited set of observables.
In particular, the progenitor radius derived using SN 2009kf ($3500~R_{\odot}$) is fully consistent with the range obtained from the independent comparison sample ($2600${--}$3900~R_{\odot}$), indicating that the inference of an extended progenitor is not driven solely by the choice of SN 2009kf as the reference object. The estimates obtained using SN 2009kf are therefore broadly consistent with those derived from other well-studied Type IIP SNe.

 The comparison sample further suggests that the explosion energy of SN 2023gfo is somewhat larger than, but still comparable to, those of ordinary Type IIP SNe. In contrast, the progenitor radius remains systematically larger than those of typical red supergiants regardless of the adopted reference object. Therefore, the inference of an unusually extended progenitor appears to be a robust result of the scaling analysis.
SN 2009kf shares several observational and inferred physical properties with SN 2023gfo, suggesting that the two SNe may have experienced a similar evolutionary process prior to explosion. \citet{Ouchi2021} interpreted the extremely large progenitor radius of SN 2009kf as evidence for an expanded envelope produced by intense pre-SN energy injection \citep{Oucih2019}. Although this interpretation is not unique, SN 2023gfo may likewise have undergone substantial pre-SN energy injection that significantly expanded its envelope before core collapse.
Although the exact physical mechanism responsible for such energy deposition remains uncertain, one possibility is energy transport by convectively excited waves generated during the final nuclear-burning stages prior to core collapse \citep[][]{Fuller2017}.


 Another possibility to explain the anomalous observational properties of SN 2023gfo is the ejecta-CSM interaction. Our first spectrum was obtained at $t = 4~\text{d}$. For some SNe IIP, the very early-phase spectrum often shows highly-ionized emission lines, such as the Balmer series as an interaction feature. However, our spectrum just shows a blue continuum. At $t = 11~\text{d}$, broad absorption lines from the ejecta component were found. This fact contradicts the interpretation of an interaction-powered SN. For interaction-powered Type IIP/IIL SNe, the spectra do not show absorption lines because the continuum is diluted by the interaction \citep[][]{Chugai2001}. 
 
 As another key observational evidence of CSM interaction, a rapid rise and bumpy structures are expected in the bluer-band light curves \citep{Moriya2024}. 
 Such features were not seen in SN 2023gfo.
Color evolution is consistent with that of other normal SNe IIP during the plateau phase (see Figure 3). Therefore, we conclude that there is no indication of ejecta-CSM interaction in our observations.

\section{CONCLUSION} 
In the present work, we presented multiband optical and NIR light curve, and optical spectra, of SN 2023gfo. From the observed light curves, bolometric luminosity, and spectral data, we derived the plateau luminosity and duration, as well as the expansion velocity. SN 2023gfo is characterized by a high peak luminosity of $(5.9 \pm 1.5)\times10^{42}$ erg s$^{-1}$, a high expansion velocity of 13,800 km s$^{-1}$, and a normal plateau duration of approximately 108 days.

We applied these observational parameters to analytical scaling relations based on \citet{Popov1993} to estimate the physical properties of the explosion and progenitor.
The results indicate an explosion energy consistent with those of typical Type IIP SNe, a slightly lower ejecta mass, and a significantly enlarged progenitor radius.

Such a combination of parameters is qualitatively consistent with scenarios involving pre-SN energy injection into the stellar envelope, which can lead to substantial radial inflation.
A similar interpretation has been proposed for the peculiar Type IIP SN 2009kf \citep{Ouchi2021}.
In this context, SN 2023gfo may represent another example of a Type IIP SN associated with an unusually extended progenitor.


This scenario can potentially explain the large progenitor radius and high luminosity derived from the observational properties of SN 2023gfo.
Alternatively, the ejecta-CSM interaction scenario may also explain the observational features, although there is no clear observational evidence for such an interaction.
Future theoretical studies should explore how pre-SN energy injection modifies the progenitor structure and whether the resulting envelope inflation can reproduce the high luminosity, normal plateau duration, and high expansion velocity observed in SN 2023gfo. Such models would provide testable predictions for light-curve and spectral evolution that could distinguish between inflated-envelope progenitors and alternative scenarios involving ejecta-CSM interaction.



\begin{acknowledgments}
 We are grateful to graduate and undergraduate students for performing 
 the NIR observations. This work was supported by Grant-in-Aid for Scientific Research (C) 22K03676. The Kagoshima University 1m telescope is a member of the Optical and Infrared Synergetic Telescopes for Education and Research (OISTER) program funded by the MEXT of Japan. D.K.S. acknowledges the support provided by DST-JSPS under grant number DST/INT/JSPS/P363/2022. This work was supported by JSPS Bilateral Program Number JPJSBP 120227709. The spectra taken by the Seimei telescope are obtained with the KASTOR
(Kanata And Seimei Transient Observation Regime) campaign (23A-N-CT10,
PI: K.M.).The Seimei telescope at the Okayama Observatory is jointly operated by Kyoto University and the National Astronomical Observatory of Japan (NAOJ),
with assistance provided by the Optical and Infrared Synergetic Telescopes for Education and Research (OISTER) program. 
K.M. acknowledges support from the Japan Society for the Promotion of Science (JSPS) KAKENHI grant Nos. JP24KK0070 and JP24H01810.
 The GROWTH-India Telescope (GIT) is a 70-cm telescope
with a 0$^{\circ}$. 7 field of view, set up by the 
 Indian Institute of
 Astrophysics (IIA) and the Indian Institute of Technology
Bombay (IITB) with funding from Indo-US Science and
Technology Forum and the Science and Engineering Research
Board, Department of Science and Technology, Government of
India. It is located at the Indian Astronomical Observatory
(IAO, Hanle). We acknowledge funding by the IITB alumni
batch of 1994, which partially supports the operation of the
telescope. Telescope technical details are available at
GROWTH-India website https://sites.google.com/view/growthindia/.
\end{acknowledgments}

\bibliography{sample631}{}
\bibliographystyle{aasjournal}

\begin{deluxetable}{ccccc}
\tabletypesize{\scriptsize}
\tablecaption{Optical photometry of SN 2023gfo obtained with the GIT\label{tab:photometry}}
\tablehead{
\colhead{JD $-$ 2400000} &
\colhead{$g$ (mag)} &
\colhead{$r$ (mag)} &
\colhead{$i$ (mag)} &
\colhead{$z$ (mag)}
} 
\startdata
60056.37 & \nodata & \nodata & \nodata & 15.327  ±  0.063 \\
60063.32 & 15.949  ±  0.026 & 15.280  ±  0.035 & 15.014  ±0.045 & 14.894±0.047 \\
60064.35 & 15.881  ±  0.026 & 15.008  ±  0.032 & \nodata & \nodata \\
60082.25 & 16.283  ±  0.044 & 15.288  ±  0.046 & 14.990 ±  0.053 & 14.817  ±  0.050 \\
60086.26 & 16.312  ±  0.021 & 15.299  ±  0.030 & 15.019  ±  0.039 & 14.772 ±  0.074 \\
60091.26 & 17.347  ±  0.278 & 15.373  ±  0.087 & 15.023  ±  0.040 & \nodata \\
60092.21 & 16.456  ±  0.055 & 15.423  ±  0.058 & 15.104  ±  0.091 & 14.855  ±  0.129 \\
60099.23 & 16.572  ±  0.063 & 15.306  ±  0.039 & 14.932  ±  0.042 & 14.866  ±  0.058 \\
60102.25 & 16.731  ±  0.026 & 15.501  ±  0.030 & \nodata & \nodata \\
60109.22 & 16.765  ±  0.054 & 15.255  ±  0.033 & 15.201  ±  0.130 & \nodata \\

\enddata
\tablecomments{Magnitudes are reported in the AB system and have not been corrected for host-galaxy reddening. Unavailable measurements are indicated by \nodata.}
\end{deluxetable}
\begin{deluxetable}{ccccc}
\tablecaption{BVRI photometry of SN 2023gfo obtained with HOWPol}
\tablehead{
\colhead{MJD} &
\colhead{$B$} &
\colhead{$V$} &
\colhead{$R$} &
\colhead{$I$}
}
\startdata
60055.78087 & $16.433\pm0.007$ & $15.707\pm0.017$ & $15.015\pm0.000$ & $14.775\pm0.001$ \\
60056.72727 & $16.406\pm0.004$ & $15.620\pm0.001$ & $15.120\pm0.002$ & $14.705\pm0.001$ \\
60060.66319 & $16.444\pm0.016$ & $15.547\pm0.005$ & \nodata & $14.607\pm0.001$ \\
60073.58303 & $16.612\pm0.003$ & $15.584\pm0.001$ & $14.885\pm0.001$ & $14.428\pm0.001$ \\
60079.64774 & $16.832\pm0.005$ & $15.619\pm0.001$ & $14.869\pm0.000$ & $14.421\pm0.003$ \\
60111.00000 & $17.990\pm0.046$ & $16.181\pm0.007$ & $15.289\pm0.002$ & $14.719\pm0.002$ \\

\enddata
\tablecomments{All magnitudes are reported in the Vega system and have not been corrected for host-galaxy reddening. Unavailable measurements are indicated by \nodata.}
\end{deluxetable}
\begin{deluxetable}{cccc}
\tablecaption{Near-infrared photometry of SN 2023gfo obtained with kSIRIUS }
\tablehead{
\colhead{MJD} &
\colhead{$J$ (mag)} &
\colhead{$H$ (mag)} &
\colhead{$K$ (mag)}
}
\startdata
60056.7 & $14.038\pm0.011$ & $13.749\pm0.020$ & $13.659\pm0.035$ \\
60060.6 & $13.954\pm0.010$ & $13.655\pm0.020$ & $13.394\pm0.021$ \\
60065.6 & $13.809\pm0.010$ & $13.245\pm0.025$ & $13.148\pm0.013$ \\
60066.6 & $13.781\pm0.009$ & $13.440\pm0.015$ & $13.168\pm0.017$ \\
60073.6 & $13.582\pm0.025$ & $13.363\pm0.046$ & $13.042\pm0.159$ \\
60079.6 & $13.617\pm0.011$ & $13.126\pm0.028$ & $12.971\pm0.021$ \\
60098.6 & $13.634\pm0.010$ & $13.134\pm0.021$ & $12.979\pm0.015$ \\
60151.5 & $14.279\pm0.025$ & $13.878\pm0.039$ & $13.510\pm0.063$ \\
60154.5 & $14.377\pm0.023$ & $14.066\pm0.029$ & $13.865\pm0.040$ \\
60304.8 & $16.934\pm0.075$ & \nodata & \nodata \\
60305.9 & $17.044\pm0.095$ & $16.162\pm0.065$ & $15.804\pm0.064$ \\
60316.8 & $17.012\pm0.068$ & $16.531\pm0.080$ & $15.966\pm0.070$ \\
60337.9 & $17.143\pm0.093$ & $16.575\pm0.099$ & $16.111\pm0.106$ \\

\enddata
\tablecomments{All magnitudes are reported in the Vega system and have not been corrected for host-galaxy reddening. Unavailable measurements are indicated by \nodata.}
\end{deluxetable}
\end{document}